\documentclass[
    ,final            
  ]
  {aipproc}

\layoutstyle{6x9}
\graphicspath{{fig/}}

\begin{document}

\title{Equation of state of nuclear matter in the first order phase transition}


\author{Toshiki Maruyama$^a$, Toshitaka Tatsumi$^b$, Satoshi Chiba$^a$}
{
address={$^a$ Advanced Science Research Center, Japan Atomic Energy Agency, Shirakata Shirane 2-4, Tokai, Ibaraki, 319-1195 Japan\\
$^b$ Department of Physics, Kyoto University, Kyoto, 606-8502 Japan}
}

\begin{abstract}

We investigate the properties of 
nuclear matter at the first-order phase transitions (FOPT) such as 
liquid-gas phase transition, kaon condensation, and hadron-quark phase transition.
As a general feature of the FOPT of matter consisting of many species of
charged particles,
there appears a mixed phases with 
regular structures called ``pasta'' due to the balance of the Coulomb repulsion
and the surface tension between two phases.
%
The equation of state (EOS) of mixed phase is different from those obtained 
by a bulk application of Gibbs conditions or by the Maxwell construction
due to the effects of the non-uniform structure.
We show that the charge screening and strong surface tension make
the EOS close to that of the Maxwell construction.
\end{abstract}

\maketitle


\section{Introduction}

Matter in compact stars has 
a wide range of density and a variety of chemical components. 
At the crust region of neutron stars,
there exists a region where the density is lower than the normal nuclear
density $\rho_0\simeq 0.16$fm$^{-3}$ over a couple of hundreds meters.
The pressure of such matter is provided by degenerate electrons,
since baryons are clustered into nuclei to form the Coulomb lattice 
and have little contribution to the pressure.
In the inner region,
pressure becomes high and density increases 
up to several times $\rho_0$.
%
Charge-neutral matter mainly consists of neutrons and the equal number of protons
and electrons under chemical equilibrium.
Since the kinetic energy of degenerate electrons is much higher than
that of baryons, the electron fraction (or the proton one) decreases
with increase of density and thus neutrons become the main component
and drip out of the nuclei.
In this way baryons as well as electrons come to contribute to the pressure.
At a certain density, other components such as hyperons and 
mesons may emerge.
For example, negative kaon condensation, expected to be of a first-order
phase transition (FOPT), remarkably softens the EOS of matter.
At even higher density, hadron-quark deconfinement transition may occur
and quarks in hadrons are liberated.
This phase transition is also considered to be of first-order.

A FOPT brings about a thermodynamic instability of uniform matter to
have phase separation.
In other words, matter should have the nonuniform mixed phase (MP) around
the critical density.
Since nuclear matter consists of two chemically independent components, 
i.e.\ baryons and electrons,
the  equalities of both baryon and electron chemical potentials 
between two phases are required by the Gibbs conditions in the MP.
Therefore the EOS of the MP cannot be obtained simply by 
the Maxwell construction, which is valid only for single component.
The components in nuclear matter are electrically charged, 
so that the local charge
neutrality is no more held in the MP.
%
To minimize the surface energy plus the Coulomb energy, which are called as the finite-size effects, 
matter is expected to form a structured mixed phase (SMP), 
i.e.\ a lattice made of lumps of one phase with a geometrical symmetry 
embedded in the other phase. The Coulomb interaction plays a minor role 
in determining the relative volume ratio of two phases, but it becomes 
important to determine the size of the structure. So we must carefully treat it in the study of the MP by taking into account the charge screening effect.

At very low densities, nuclei in matter are expected to form the
Coulomb lattice embedded in the electron sea.
With increase of density, ``nuclear pasta'' structures 
emerge as a SMP \cite{Rav83} in the liquid-gas phase transition, 
where stable nuclear shape may change from droplet to rod, slab, tube, and to bubble. 
Pasta nuclei are eventually dissolved into uniform matter at a certain nucleon 
density below the saturation density $\rho_0$. 
The name ``pasta'' comes from rod and slab structures 
figuratively spoken as ``spaghetti'' and ``lasagna''.
Such low-density nuclear matter exists in the collapsing stage of
supernovae and in the crust of neutron stars.
%
The SMP is also expected in the phase transitions at
higher densities, 
like kaon condensation and 
hadron-quark phase transition.
There, the charge screening effect may be pronounced 
since the local charge density can be high.

Our purpose here is 
to investigate 
pasta structures of MP
self-consistently within  
the mean-field approximation. 
In particular, 
we figure out how the Coulomb screening and the surface tension
affect the property of the MP.

\section{Relativistic mean field calculation of 
low-density nuclear matter at zero temperature}

First we investigate the property of low-density nuclear matter
at zero temperature.
We employ the relativistic mean field (RMF) model to describe the
properties of nuclear matter under consideration.
The RMF model with fields of mesons and baryons introduced
in a Lorentz-invariant way is not only relatively simple for 
numerical calculations, but also sufficiently
realistic to reproduce bulk properties of finite nuclei
as well as the saturation properties of nuclear matter \cite{maru05,marurev}.
One characteristics of our framework is that
the Coulomb interaction is properly included in the 
equations of motion for nucleons and electrons and for meson mean fields.
Thus the baryon and electron density profiles, as well as the meson
mean fields, are determined in a fully
consistent way with the Coulomb interaction.

To solve the equations of motion for the fields numerically,
we divided the whole space into equivalent Wigner-Seitz cells with 
geometrical symmetry. 
The shape of the cell changes as
sphere in three-dimensional (3D) case, cylinder in 2D and slab in 1D.
Each cell is globally charge-neutral and all physical quantities
in the cell are smoothly connected to those of the next cell
with zero gradients at the boundary.
The coupled equations for fields in a cell are solved by a relaxation method
for a given baryon-number density 
under constraints of the global charge neutrality.
Parameters included in the RMF model are chosen to reproduce the saturation properties
of symmetric nuclear matter, 
and
the binding energies and the proton ratios of nuclei.
Details of the numerical procedure are explained in Refs.\ \cite{maru05,marurev}.

\section{Pasta structure and the equation of state}

\begin{figure*}
\includegraphics[width=.89\textwidth]{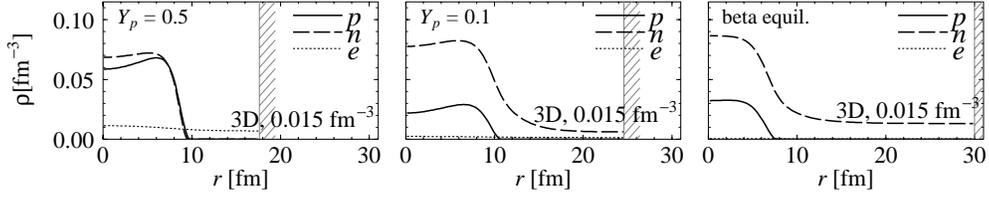}
\caption{
Examples of the density profiles in the 3D cells for symmetric nuclear
 matter with $Y_p$=0.5 (left), matter
 with $Y_p=0.1$ (center) and matter in beta-equilibrium (right).
}
\label{proffixfull}
\end{figure*}

Here, we present our results on
nuclear matter with a fixed proton fraction $Y_p=0.5$ and $0.1$,
and nuclear matter in beta-equilibrium.
Figure \ref{proffixfull} shows
some typical density profiles inside the Wigner-Seitz cells
with the 3D geometry. 
The horizontal axis in each panel
denotes the radial distance from the center of the cell. 
The cell boundary is indicated by the hatch.
These density distributions exhibit nuclear ``droplets''  
among five kinds of ``pasta'' structures.
In symmetric matter ($Y_p =0.5$), two phases consist of
nuclear liquid and electron gas.
This is the specific case for $T=0$, 
since both the phases include electron and nucleons at finite temperature. 
For the cases of $Y_p =0.1$ and beta-equilibrium, 
the neutron density is finite at any point:
two phases consist of nuclear matter and dripped neutrons, besides electrons.
%
%

Due to the spatial rearrangement of electrons the electron
density profile becomes inhomogeneous. 
This non-uniformity of the electron distribution is
more pronounced for a higher $Y_p$ and a higher density. 
Protons repel each other.
Thereby the proton density profile substantially deviates from the 
uniform distribution.

\begin{figure*}
\includegraphics[width=.31\textwidth]{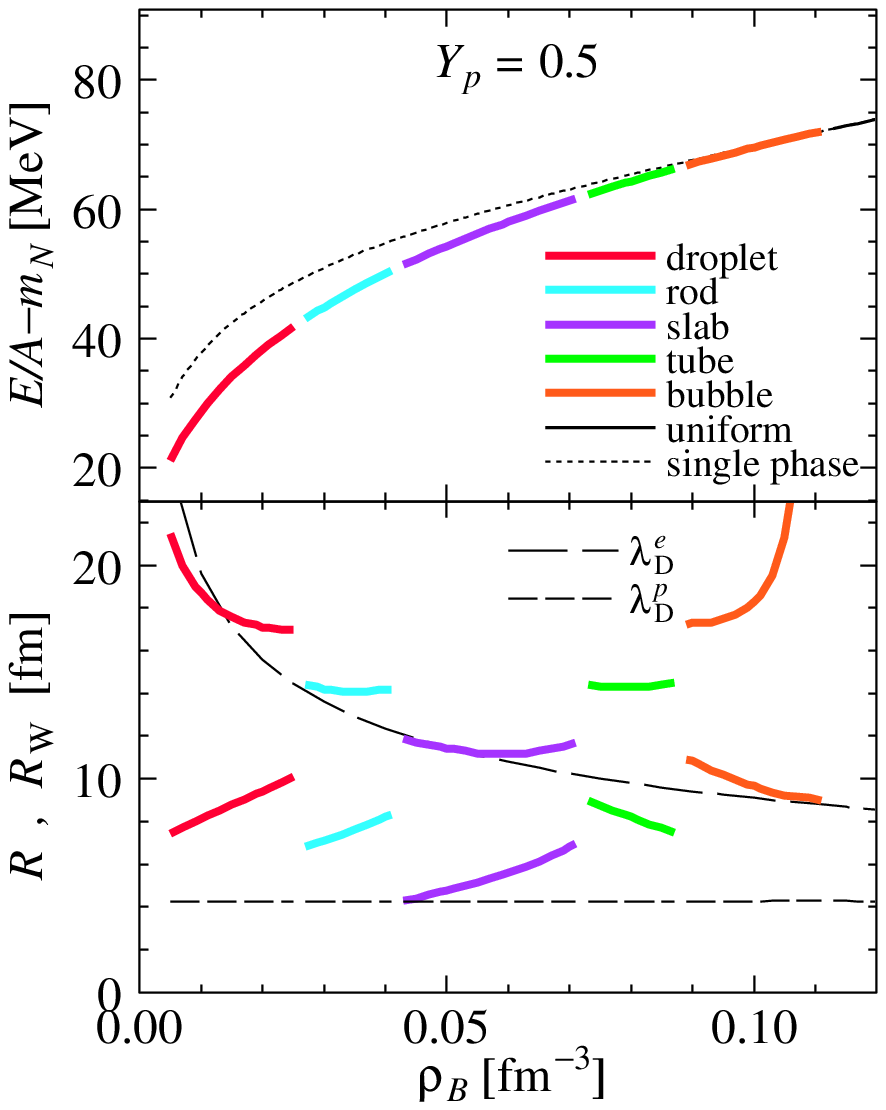}
\includegraphics[width=.31\textwidth]{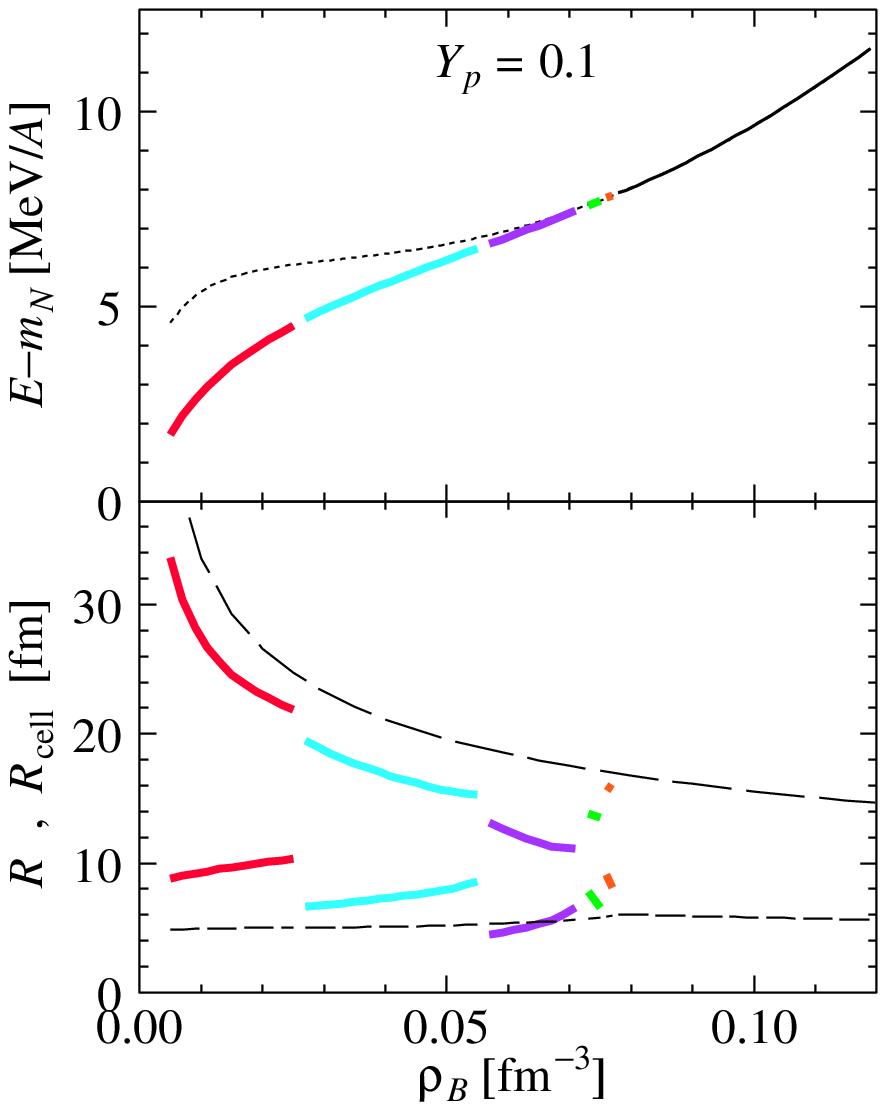}
\includegraphics[width=.31\textwidth]{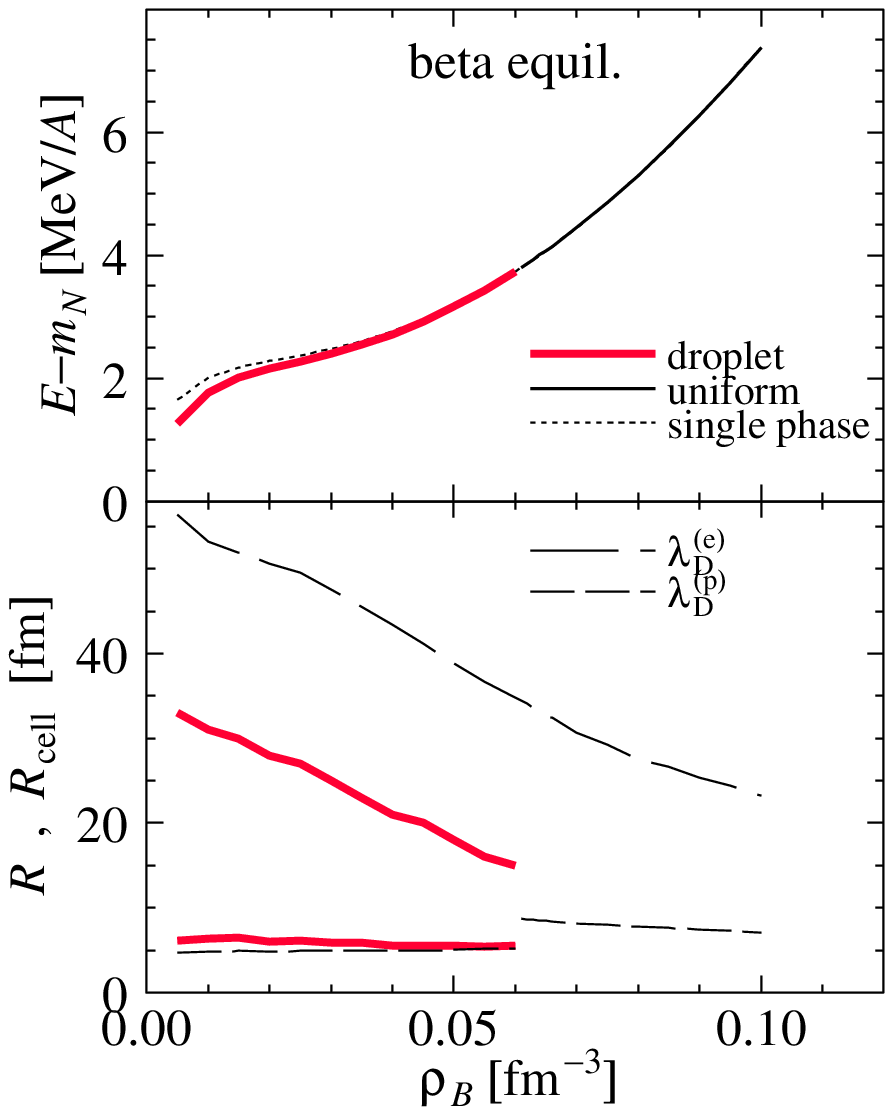}
\caption{
Binding energy per nucleon (upper) and the cell and lump sizes (lower) 
for symmetric nuclear matter with $Y_p$=0.5 (left), matter with
$Y_p=0.1$ (center), and matter in beta-equilibrium (right).
}
\label{eosfixfull}
\end{figure*}

The EOS with the sequence of the geometric structures is
shown in Fig.~\ref{eosfixfull} (upper panels)
as a function of the averaged baryon-number density.
Note that the energy $E/A-m_N$ also includes the
kinetic energy of electrons, which makes the total pressure positive. 
The lowest-energy configurations are selected among
various geometrical structures. 
In the cases of fixed proton fraction, 
the most favorable configuration
changes from the droplet to rod, slab, tube, bubble, and to the
uniform one (the dotted thin curve) with an increase of density.
The appearance of non-uniform structures in matter results in a
softening of EOS: the energy per baryon
gets lower up to about 15 MeV$/A$ compared to uniform matter.
In the case of beta-equilibrium, on the other hand,
only the droplet configuration is seen for the MP.
However, this is dependent on the interaction 
used in the calculation \cite{oyamatsu}. 
The energy gain due to the appearance of non-uniform matter
in beta-equilibrium case is rather small.

\section{Clustering mechanism}

\begin{figure*}
\includegraphics[width=.37\textwidth]{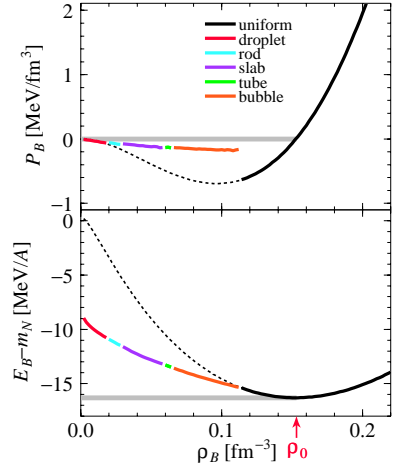}
\caption{
Baryon partial pressure (upper) and baryon binding energy (lower) 
for matter with $Y_p=0.5$. \vspace{-2mm}
}
\label{figPartialPressure}
\end{figure*}

Let us explain the mechanism of clustering (appearance of non-uniform matter).
The upper panel of Fig.~\ref{figPartialPressure} shows baryon partial pressure.
The dotted curve and thin full curve shows the case of uniform matter.
Non-uniform structure (pasta) appears in the density region where uniform
matter becomes unstable with negative partial pressure $P_B$.
Then the resultant $P_B$ increases up to near zero.

One should note there still remains a region where uniform
matter has negative $P_B$.
This is due to the finite-size effects (surface tension and 
the Coulomb potential) which make the pasta structure unstable.
Consider 3D case for simplicity.
The Coulomb energy par particle is proportional to 
second power of the droplet size, $E_C\propto R^2$,
and the surface energy per particle 
is proportional to its inverse, $E_{\rm surf}\propto R^{-1}$.
Then the sum $E_C+E_{\rm surf}$ has a
minimum at a certain $R$ with $2E_C=E_{\rm surf}$.
However, the charge screening reduces the Coulomb energy.
This moves the minimum point to larger $R$.
At some density near $\rho_0$, the screening effect is
strong and $E_C$ is reduced significantly.
This extinguishes the minimum point and makes the pasta 
structure unstable.

\section{Low-density nuclear matter at finite temperature}

Next we explore the case of finite temperature $T>0$.
For this we extend our framework of the previous section.
For simplicity only the distribution functions of Fermions
are treated as $T$-dependent; the momentum distribution function is
a Fermi-Dirac distribution instead of a step function with a threshold 
at the Fermi momentum.
In the numerical calculation,
density, scalar density, and kinetic energy density, etc of a fermion $a$ 
are obtained by integrating the functions of $T$, $\mu_a$ and $m_a^*$
over all the momentum-space.
We store those values in tables and get necessary quantities by
interpolating them.
The contribution of anti-particles are neglected for simplicity.

\begin{figure*}[t]
\includegraphics[width=.48\textwidth]{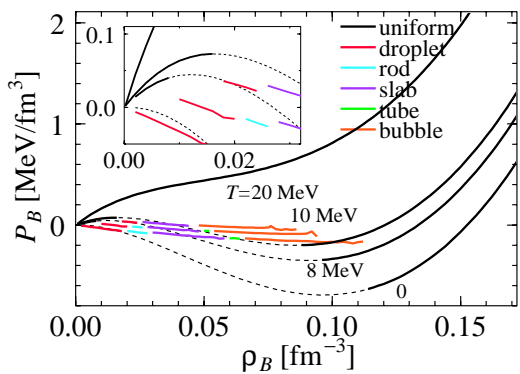}
\caption{
Baryon partial pressure (upper) 
for symmetric nuclear matter 
at various temperatures.
The inner box shows a enlargement of 
the low-density region.
}
\label{figPartialPressureT}
\end{figure*}

We have observed ``pasta'' structures again at finite but 
low temperatures.
The difference from the $T=0$ case is that the low-density phase
always contains some amount of protons and neutrons. 
The resultant EOS (baryon partial pressure as a function of baryon number density)
of symmetric nuclear matter 
at various temperatures is shown in Fig.~\ref{figPartialPressureT}.
Dotted and thick solid curves show the cases of uniform matter,
while thin solid curves are the cases where non-uniform structures
are present.
Pasta structures appear at finite temperatures as well as
the case of $T=0$.
But there appears uniform matter (gas phase) at the lowest-density region
\cite{avancini,friedman}
since the baryon partial pressure of uniform matter has
a positive gradient against density.
On the other hand, the uniform matter is unstable
where the pressure gradient is negative
even if the pressure itself is positive.
At $T=20$ MeV, we obtain no pasta structure
since the baryon partial pressure of uniform matter
is a monotonic function of density.
So the critical temperature lies between 10 and 20 MeV.
Study of the (spinodal) instability of uniform matter 
is in progress. 

\section{High density nuclear matter}

In the core region of neutron stars,
first-order phase transitions such as
meson condensation and quark deconfinement transition, etc
are expected.
We have investigated the similar pasta structures
in such FOPTs.
Particularly the hadron-quark MP is a good 
example to show the effects of charge-screening.

\begin{figure*}
\includegraphics[width=.48\textwidth]{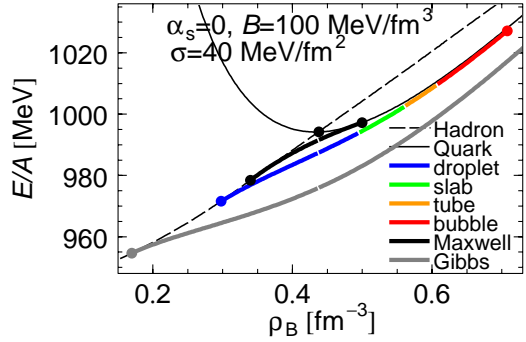}
\caption{
EOS of matter at the hadron-quark phase transition.
Solid and dashed curves at higher energies indicate
quark and hadron single phases, respectively.
Gray curves indicate EOS of MP. 
}
\label{figEOS-QH}
\end{figure*}

Figure \ref{figEOS-QH} shows the EOS of matter
at hadron-quark phase transition.
The appearance of MP significantly lower
the energy of the system.
The EOS obtained by our calculation with
pasta structures is close to that obtained by
the Maxwell construction shown by a thick black curve.
This is due to the strong surface tension between two phases
and the charge screening which makes the system approximately
local-neutral.
This situation is similar to that in the Maxwell construction.
On the other hand, if the surface tension is weak,
the size of the structure becomes small \cite{marukaon,maruQH}
and consequently the Coulomb interaction becomes ineffective.
This situation corresponds to a bulk application of the Gibbs
conditions without the finite-size effects.

In conclusion we emphasize that the existence of
pasta structures together with the above mentioned 
surface and the charge screening effects are
common and general for the MPs at 
the FOPT in nuclear matter.  


\begin{theacknowledgments}
This work has been done by the collaborations with
T.~Tanigawa, D.~N.~Voskresensky, T.~Endo, 
H.-J.~Schulze and Tomoyuki Maruyama.
\end{theacknowledgments}



\bibliographystyle{aipproc}   


\end{document}